\documentstyle[12pt]{article}
\begin{document}

\def \d{{\rm d}}
\newcommand{\pul}{\frac{1}{2}}
\newcommand{\la}{c}

\title{Non-expanding impulsive gravitational waves}

\author{J. Podolsk\'y\thanks{E--mail: {\tt podolsky@mbox.troja.mff.cuni.cz}}
\\
\\ Department of Theoretical Physics,\\
Faculty of Mathematics and Physics, Charles University,\\
V Hole\v{s}ovi\v{c}k\'ach 2, 180~00 Prague 8, Czech Republic.\\}

\maketitle

\begin{abstract}
We investigate a class of impulsive gravitational waves which propagate
either in Minkowski or in the (anti--)~de~Sitter background. These waves
are constructed as impulsive members of the Kundt class $P(\Lambda)$ of
non-twisting, non-expanding  type {\it N} solutions of vacuum Einstein
equations with a cosmological constant $\Lambda$.

We show that the only non-trivial waves of this type in Minkowski spacetime
are impulsive {\it pp}-waves. For $\Lambda\not=0$ we demonstrate that
the canonical subclasses of $P(\Lambda)$, which are invariantly different
for smooth profiles, are all locally equivalent for impulsive profiles.

Also, we present coordinate system for these
impulsive solutions which is explicitly continuous.

\end{abstract}

\vfil\noindent
PACS class: 04.20.Jb, 04.30.Nk, 04.30.-w, 98.80.Hw

\bigskip\noindent
Keywords: impulsive waves, cosmological constant

\newpage
\section {Introduction}
\medskip\noindent
In a now classic paper \cite{Penrose}, Penrose proposed a method
for constructing general impulsive {\it pp}-waves
(plane-fronted gravitational waves with parallel rays) and
spherical waves of the Robinson-Trautman type \cite{KSMH80}.
His ``scissors-and-paste'' approach is based on the removal of a
null hypersurface from Minkowski space-time and re-attaching the
parts by making an identification with a ``warp''. As an explicit
example Penrose constructed an impulsive plane wave. A family of
impulsive spherical gravitational waves obtained by the above
method have later been investigated by Hogan
\cite{Hogan92}-\cite{Hogan94}.

Another explicit impulsive {\it pp}-wave solution was found by
Aichelburg and Sexl \cite{AicSex71}. However, it was obtained
by a different method: by boosting a Schwarzschild black hole
to the speed of light while its mass is reduced to zero
in an appropriate way.  This interesting solution describes
an impulsive wave generated by a single null particle.
More general metrics have similarly been obtained by boosting
other space-times of the Kerr-Newman family
\cite{FerPen}-\cite{BaNa96}.

A similar approach has been adopted by Hotta and Tanaka \cite{HotTan93}
in boosting the Schwarz\-schild--(anti-)de~Sitter metric to a similar limit.
This approach also results in impulsive gravitational waves.
However, the background in this case is not flat and the wavefronts are not
simple planes as in Minkowski space. In \cite{PodGri97} we demonstrated
that the impulsive wave surface in the de~Sitter background is a
non-expanding sphere forming the cosmological horizon. It is generated
by two null particles located at the poles. In the anti-de~Sitter
background the wavefronts are hyperboloidal surfaces
containing a single null particle which propagates from one side
of the universe to the other and then returns in an endless cycle.
We also determined the global structure and conformal diagrams for
these spacetimes.

In subsequent papers \cite{GriPod97}, \cite{PodGri98} we presented
a more general class of such solutions. Their global structure and
the shape of the impulsive wave surfaces are exactly as summarized
above but the sources are null particles which can have an
arbitrary multipole structure and which are distributed arbitrarily
over the wave surface. We also noticed that these
space-times are related to the impulsive limit of a family
of non-twisting and non-expanding type~N solutions of the Kundt class
studied in \cite{ORR85}. It is the purpose of the present paper to
investigate the relation systematically.

In section~2 we summarize properties of the Kundt class $P(\Lambda)$
and, in particular, we present its canonical subclasses. Impulsive waves
of this type for $\Lambda=0$ are investigated in section~3.
Sections~4 and~5 are devoted to non-expanding impulsive waves
in de~Sitter and anti--de~Sitter backgrounds. Continuous coordinate
systems for all these  solutions are presented in the final section~6.

\section {Non-expanding gravitational waves $P(\Lambda)$}
\medskip\noindent
It can be shown \cite{BP} that all non-twisting type {\it N} solutions
of the vacuum Einstein field equations with (possibly) non-vanishing cosmological
constant $\Lambda$ belong either to the Kundt class of non-expanding
gravitational fields or to the expanding Robinson-Trautman class.
In this paper we concentrate on the Kundt class which we denote $P(\Lambda)$.

This class has been investigated by Ozsv\' ath, Robinson
and R\' ozga \cite{ORR85} and in \cite{BP}.
In a suitable coordinate system where $v$ is a parameter along
null geodesics generated by the quadruple Debever-Penrose
vector field, $u$ is a retarded time with $u=$const. being
wavefronts and $\xi,\bar\xi$ are space-like coordinates,
all the $P(\Lambda)$ solutions can be written as
\begin{equation}
\d s^2=2\frac{q^2}{p^2}\d u\d v-2\frac{1}{p^2}\d\xi \d\bar\xi - F \d u^2\ ,
\label{E2.2}
\end{equation}
where
\begin{eqnarray}
&&p=1+ \la \xi\bar\xi\ , \nonumber \\
&&q=(1-\la \xi\bar\xi)\alpha+\bar\beta\xi+\beta\bar\xi\ , \nonumber \\
&&F=\kappa{q^2\over p^2}v^2 - 2l{q^2\over p^2}v + {q\over p}H\ , \nonumber \\
&&c={\Lambda \over 6}\ , \qquad
  \kappa=2c\alpha^2+2\beta\bar\beta\ ,  \qquad
  l={(\ln |q|)},_u \ . \nonumber
\end{eqnarray}
Here $\alpha(u)$ and $\beta(u)$ are arbitrary real and complex functions of
$u$, respectively. These functions play the role of two additional parameters,
i.e. $P(\Lambda)\equiv P(\Lambda)[\alpha,\beta]$.
Finally, $H=H(\xi,\bar\xi,u)$ is a function restricted by
the vacuum field equations with a cosmological constant,
$H_{,\xi\bar\xi}+2cH/p^2=0$.
A general solution to this equation is
\begin{equation}
H(\xi,\bar\xi,u)=(f_{,\xi}+\bar f_{,\bar\xi})-{2c\over p}
    (\bar\xi f + \xi\bar f)\ ,
\label{E2.5}
\end{equation}
where $f(\xi,u)$ is an arbitrary function of $\xi$ and $u$, analytic in
$\xi$. The space-time is conformally flat
(and thus of a constant curvature)
if and only if the function $H$ is of the form
$H=H_c={\Big[(1-\la \xi\bar\xi){\cal A}+
         \bar {\cal B}\xi+{\cal B}\bar\xi\Big]}/p$
with ${\cal A}(u)$ and ${\cal B}(u)$ being arbitrary real and complex functions, respectively.
Since $H_c$ of this form corresponds to (\ref{E2.5}) for $f$ quadratic in $\xi$
we infer that the $P(\Lambda)$ solutions (\ref{E2.2}), (\ref{E2.5}) with
$f=f_c\equiv c_0(u)+c_1(u)\xi+c_2(u)\xi^2$, where $c_i(u)$ are arbitrary
complex functions of $u$, are isometric to Minkowski, de Sitter
and anti-de Sitter space-time if $\Lambda=0$, $\Lambda>0$ and
$\Lambda<0$, respectively.

As was suggested in \cite{ORR85}, one can base the invariant
`canonical' classification of all the $P(\Lambda)$ solutions
on  signs of the parameters $\kappa$ and  $\Lambda$ (see also \cite{BP}).
There are nine  possible cases since $\kappa$ and $\Lambda$ can
both be positive, zero or negative. However, subclasses $\kappa<0 , \Lambda>0$
and $\kappa=0 , \Lambda>0$  and  $\kappa<0 , \Lambda=0$
are forbidden since they violate the relation
$\kappa=2c\alpha^2+2\beta\bar\beta$. The remaining six possibilities
give the following subclasses:

\medskip
\noindent
{\bf A. Subclass $\kappa=0 , \Lambda=0$.}
Using a coordinate freedom, a canonical representative of the subclass
can be found and denoted as

 \centerline{$PP\equiv P(\Lambda=0)[\alpha=1,\beta=0]$\ .}

\noindent
 We are  using the symbol  $PP$  since the corresponding metric
\begin{equation}
\d s^2=2\d u\d v -2d\xi d\bar\xi - 2(g+\bar g)\d u^2\ ,
\label{E2.14}
\end{equation}
 with $g(\xi,u)=\pul f_{,\xi}$  analytic in $\xi$, describes the well-known
 plane-fronted gravitational waves with parallel rays ({\it pp}-waves).
 These solutions have thoroughly been investigated by many authors;
 for details  and references  see \cite{KSMH80}.

\medskip
\noindent
{\bf B. Subclass $\kappa>0 , \Lambda=0$.}
 Any solution of this subclass can be put into the form

 \centerline{$PK\equiv P(\Lambda=0)[\alpha=0,\beta=1]$\ .}

\noindent
 We have used the symbol $PK$ since the metric
\begin{equation}
\d s^2= 2(\xi+\bar\xi)^2 \d u \d v - 2 \d\xi\d\bar\xi -2(\xi+\bar\xi)^2
 \left(v^2+\frac{g+\bar g}{\xi+\bar\xi}\right)\d u^2\ ,
\label{E2.15}
\end{equation}
 with $g(\xi,u)=\pul f_{,\xi}$ describes the wave solutions with $\Lambda=0$
 discovered by Kundt \cite{Kundt61}.  Details about this solution can
 also be found  in \cite{KSMH80}.

\bigskip
\noindent
{\bf C. Subclass $\kappa>0 , \Lambda>0$.}
We can always find a transformation to the canonical representative
of this subclass

 \centerline{$PK(\Lambda)I\equiv P(\Lambda)[\alpha=0,\beta=1]$\ .}

\noindent
The corresponding metric reads
\begin{equation}
\d s^2=2{\left({{\xi+\bar\xi}\over{1+\la\xi\bar\xi}}\right)}^2\d u\d v
    -2{{\d\xi\d\bar\xi}\over{{(1+\la\xi\bar\xi)}^2}}
    -\left[ { 2{\left({{\xi+\bar\xi}\over{1+\la\xi\bar\xi}}\right)}^2 v^2
      + {{\xi+\bar\xi}\over{1+\la\xi\bar\xi}}H} \right]\d u^2
 \ . \label{E2.16}
\end{equation}
 We have introduced  $PK(\Lambda)I$ in order to indicate that this solution
 represents a generalization of the Kundt waves $PK$ to the case
 $\Lambda\not=0$.  Here `I' means `of the first kind' since there are two
 other $PK(\Lambda)$ solutions. The solution $PK(\Lambda)I$
 was first discovered in \cite{GarPle81}.

\bigskip
\noindent
{\bf D. Subclass $\kappa>0 , \Lambda<0$.}
 As in the previous case the representative is
 $PK(\Lambda)I=P(\Lambda)[\alpha=0,\beta=1]$. The  metric
 (\ref{E2.16}) has the same form for negative cosmological constant
 $\Lambda=6c<0$.

\bigskip
\noindent
{\bf E. Subclass $\kappa<0 , \Lambda<0$.}
 The canonical form of this subclass is

\centerline{$PK(\Lambda^-)II\equiv P(\Lambda<0)[\alpha=1,\beta=0]$}

\noindent
and the metric has the form
\begin{equation}
\d s^2=2{\left({{1-\la\xi\bar\xi}\over{1+\la\xi\bar\xi}}\right)}^2\d u\d v
   -2{{\d\xi\d\bar\xi}\over{{(1+\la\xi\bar\xi)}^2}}
   -\left[{ 2c{\left({{1-\la\xi\bar\xi}\over{1+\la\xi\bar\xi}}\right)}^2 v^2
    +{{1-\la\xi\bar\xi}\over{1+\la\xi\bar\xi}}H} \right]\d u^2\ .
\label{E2.17}
\end{equation}
 The symbol $PK(\Lambda^-)II$ means generalized Kundt waves
 `of the second kind' with a negative cosmological constant. This class
 was first discovered in \cite{ORR85}.

\bigskip
\noindent
{\bf F. Subclass $\kappa=0 , \Lambda<0$.}
 In this case the canonical representative is
$PK(\Lambda^-)III\equiv
P(\Lambda<0)[\alpha=1,\beta=\sqrt{-\la}\,e^{i\omega(u)}]$,
$\omega(u)$ being an arbitrary real function of $u$.
In particular, if the Debever-Penrose vector is a Killing vector,
$\beta=$const. and the representative is

 \centerline{$PK(\Lambda^-)III_K\equiv
               P(\Lambda<0)[\alpha=1,\beta=\sqrt{-\la}\,]$\ ,}
\noindent
where the suffix `K' stands for `Killing'. The metric in this
case reads
\begin{equation}
\d s^2=2{\left({{1+\sqrt{-\la}\,(\xi+\bar\xi)-c\xi\bar\xi}
             \over{1+\la\xi\bar\xi}}\right)}^2 \d u\d v
 -2{{\d\xi\d\bar\xi}\over{{(1+\la\xi\bar\xi)}^2}}
 -{{1+\sqrt{-\la}\,(\xi+\bar\xi)-c\xi\bar\xi}\over{1+\la\xi\bar\xi}}H\d u^2
\  .\label{E2.18}
\end{equation}
It can be shown \cite{Podol98} that this subclass is the same as
the `Lobatchevski waves' studied by Siklos \cite{Siklos85}.
\bigskip

Finally, we summarize the invariant classification of the
$P(\Lambda)$ class of solutions in the following diagram:
\[ P(\Lambda)\ \left\{\begin{array}{ll}
            \ \Lambda=0\      \left\{\begin{array}{l}
                       \kappa=0: PP  \\
                       \kappa>0: PK  \\
                       \end{array}\right. \\
                       &\\
            \ \Lambda\not=0\  \left\{\begin{array}{l}
                       \kappa>0: PK(\Lambda)I  \\
                       \kappa<0: PK(\Lambda^-)II  \\
                       \kappa=0: PK(\Lambda^-)III \rightarrow
                                 PK(\Lambda^-)III_K \end{array}\right. \\
             \end{array}\right. \]
There are three subclasses for $\Lambda<0$ but only one subclass for $\Lambda>0$.
The reason for this asymmetry is a geometrical one,
embodied in the condition $\kappa=2c\alpha^2+2\beta\bar\beta$
which for $\Lambda>0$ excludes the cases $\kappa<0$ and $\kappa=0$.
We also observe  natural relations between
$\Lambda=0$ and $\Lambda\not=0$ subclasses. We can set
$\Lambda=6c=0$ in the metrics (\ref{E2.16}), (\ref{E2.17}),
(\ref{E2.18}) and find that
$PK(\Lambda=0)I=PK$, $PK(\Lambda=0)II=PP=PK(\Lambda=0)III_K$.
Thus, it is natural to consider the $PK(\Lambda)I$ class as a
generalization of the Kundt solution $PK$ whereas the
classes $PK(\Lambda^-)II$  and $PK(\Lambda^-)III_K$ as generalizations of
{\it pp}-waves.

\section {Non-expanding impulsive waves with $\Lambda=0$}
\medskip\noindent
From the above classification of the $P(\Lambda)$ class of
non-expanding gravitational waves we observe that there are two
subclasses for vanishing cosmological constant, $PP$ and $PK$.
Therefore, we can easily construct impulsive waves in Minkowski
spacetime by assuming $\delta$-function profiles of the form
$g(\xi,u)=g(\xi)\delta(u)$ in metrics
(\ref{E2.14}) or (\ref{E2.15}). These two types of impulsive
solutions look quite different but we can demonstrate that, in fact,
they are locally {\it equivalent}. Indeed, a transformation
\begin{eqnarray}
   U   &=& (\xi+\bar\xi)(1+uv)u\ , \nonumber\\
   V   &=& (\xi+\bar\xi)v      \ , \label{E3.1}\\
   \eta&=& (\xi+\bar\xi)uv+\xi \ , \nonumber
\end{eqnarray}
converts the impulsive $PK$ solution (\ref{E2.15}) to
\begin{equation}
\d s^2=2\d U\d V -2d\eta d\bar\eta
 - 2 [g(\eta)+\bar g(\bar\eta)]\delta(U)\d U^2\ , \label{E3.2}
\end{equation}
which is exactly the $PP$ impulsive solution (\ref{E2.14}). To be
more precise, we should note that an inverse transformation
to (\ref{E3.1}) is
\begin{eqnarray}
   u   &=& {1\over 2V}
     [\eta+\bar\eta\mp\sqrt{(\eta+\bar\eta)^2-4UV}\,]\ , \nonumber\\
   v   &=& \pm{V/\sqrt{(\eta+\bar\eta)^2-4UV}}\ , \label{E3.3}\\
   \xi&=& {1\over2}[\eta-\bar\eta\pm\sqrt{(\eta+\bar\eta)^2-4UV}\,]\ .\nonumber
\end{eqnarray}
One can show that $(\eta+\bar\eta)^2-4UV=(\xi+\bar\xi)^2\ge0$.
The transformation (\ref{E3.3}) is not unique but the upper
signs parametrize the part of spacetime given by $\xi+\bar\xi>0$
whereas the lower signs parametrize the part given by
$\xi+\bar\xi<0$. For $\xi+\bar\xi=0$ there is a coordinate
singularity in (\ref{E2.15}).

To summarize, the only non-trivial impulsive gravitational waves
of the form (\ref{E2.2}) in Minkowski spacetime are impulsive
{\it pp}-waves (\ref{E3.2}).

\section {Non-expanding impulsive waves with $\Lambda\not=0$}
\medskip\noindent
In \cite{PodGri98} we presented a general class of exact solutions of
Einstein's equations which describe impulsive gravitational waves in
de~Sitter or anti-de~Sitter background generated by null particles with an
arbitrary multipole structure. In five-dimensional notation
the solutions can simply be written in the form
 \begin{eqnarray}
\d s^2&=& \d{Z_0}^2 -\d{Z_1}^2 -\d{Z_2}^2 -\d{Z_3}^2
-\epsilon\d{Z_4}^2 \nonumber \\
 &&\qquad -H(Z_2,Z_3,Z_4)\delta(Z_0+Z_1)(\d Z_0+\d Z_1)^2  \ .
\label{E4.1}
 \end{eqnarray}
The de~Sitter or the anti-de~Sitter background can be represented
as a metric (\ref{E4.1}) with $H=0$  on the four-dimensional hyperboloid
 \begin{equation}
 {Z_0}^2 -{Z_1}^2 -{Z_2}^2 -{Z_3}^2 -\epsilon{Z_4}^2 =-\epsilon a^2\ , \label{E4.2}
 \end{equation}
where $a^2=3/\epsilon\Lambda$,
$\epsilon=1$ for the de~Sitter background ($\Lambda>0$), and $\epsilon=-1$ for
the anti-de~Sitter background ($\Lambda<0$).
The impulse is located on the null hypersurface given by
 \begin{equation}
 Z_0+Z_1=0, \qquad {Z_2}^2 +{Z_3}^2 +\epsilon{Z_4}^2 =\epsilon a^2,
\label{E4.3}
\end{equation}
which is a 2-sphere for $\epsilon=1$ or a 2-dimensional hyperboloid
for $\epsilon=-1$. The function $H(Z_2,Z_3,Z_4)$ is determined on the
wave surface (\ref{E4.3}), i.e., it must be a function of two parameters
which span the surface. An appropriate parametrization is thus given by
 \begin{equation}
 Z_2=a\sqrt{\epsilon(1-z^2)}\cos\phi, \qquad
Z_3=a\sqrt{\epsilon(1-z^2)}\sin\phi, \qquad Z_4=az\ , \label{E4.4}
 \end{equation}
 where $|z|\le1$ when $\epsilon=1$ and $|z|\ge1$ when $\epsilon=-1$.
We showed in \cite{PodGri98} that a non-trivial vacuum solution
of this type is given by
 \begin{equation}
 H(z,\phi)= \sum_{m=0}^\infty b_mH_m(z,\phi)
= \sum_{m=0}^\infty b_mQ^m_1(z)\cos[m(\phi-\phi_m)]\ , \label{E4.5}
\end{equation}
 where $b_m$ and $\phi_m$ are real constants representing an arbitrary
amplitude and phase of each component and $Q^m_1(z)$ are associated
Legendre functions of the second kind generated by the relation
\begin{equation}
Q^m_1(z)=(-\epsilon)^m|1-z^2|^{m/2}{\d^m\over\d z^m}Q_1(z)\ ,\label{E4.6}
\end{equation}
where $Q_1(z)=Q^0_1(z)=(z/2)\log\left|(1+z)/(1-z)\right|-1$.
The first term ($m=0$) gives the simplest (axially symmetric)
solution found by Hotta and Tanaka \cite{HotTan93}. In general,
these components describe point sources with an $m$-pole structure
\cite{PodGri98}. Geometry of the spacetimes was investigated in
\cite{PodGri97}.

\section {Equivalence of all the $P(\Lambda\not=0)$ subclasses of
          impulsive waves}
\medskip\noindent
In this section we shall demonstrate that although the invariant canonical
subclasses of the $P(\Lambda\not=0)$ class of non-expanding gravitational
waves described in section 2 are generically different for smooth
profiles, they are locally {\it equivalent for impulsive profiles}.
Moreover, they are just different four-dimensional
parametrizations of the general impulsive solution
in de Sitter or anti--de Sitter spacetimes given by Eqs.
(\ref{E4.1})--(\ref{E4.4}).

For the $PK(\Lambda>0)I$ class of impulsive solutions with
$H=H(\xi,\bar\xi)\delta(u)$
it was shown in \cite{PodGri98} that performing the transformation
\begin{equation}
 \xi=\sqrt2 a \,e^{i\phi}\tan{{\theta\over2}}, \qquad\qquad
 v={a\over\sqrt2}{1\over t}, \qquad\qquad
 u={1\over\sqrt2a}(\rho-t),  \label{E5.2}
\end{equation}
 where $a=1/\sqrt{2\epsilon c}=\sqrt{3/\Lambda}$ the line element
(\ref{E2.16}) becomes
 \begin{eqnarray}
 && \d s^2 ={a^2\over t^2}\sin^2\theta\cos^2\phi \> (\d t^2-\d\rho^2)
-a^2(\d\theta^2+\sin^2\theta\d\phi^2) \nonumber \\
 &&\qquad\qquad\qquad -\sin\theta\cos\phi \,
H(\theta,\phi)\delta(t-\rho)(\d t - \d\rho)^2.
 \label{E5.6}
 \end{eqnarray}
 This can be seen to be exactly the solution (\ref{E4.1}) in which the
de~Sitter background in the five-dimensional form (\ref{E4.2}) is
parametrized by
 \begin{eqnarray}
 Z_0 &=& -a\cos\theta +{a^2\over t}\sin\theta\cos\phi
+{\rho^2-t^2\over2t}\sin\theta\cos\phi \nonumber\\
 Z_1 &=& \quad a\cos\theta -{a^2\over t}\sin\theta\cos\phi \nonumber\\
 Z_2 &=& {\rho\over t}\>a\sin\theta\cos\phi \\
 Z_3 &=& \quad a\sin\theta\sin\phi \nonumber\\
 Z_4 &=& \quad a\cos\theta \hskip6pc
-{\rho^2-t^2\over2t}\sin\theta\cos\phi\ . \nonumber
 \label{E5.7}
 \end{eqnarray}

For $\Lambda<0$ the metric (\ref{E2.16}) of impulsive $PK(\Lambda)I$
subclass with $H=H(\xi,\bar\xi)\delta(u)$ can be expressed,
using a similar transformation
 \begin{equation}
 \xi = \sqrt{2} a\, e^{i\phi_1}\tanh{R_1\over 2},\qquad
   v = {a\over\sqrt{2}}{1\over t_1},\qquad
   u = {1\over \sqrt{2}a} (\rho_1-t_1),  \label{E5.8}
 \end{equation}
where $a=1/\sqrt{-2\epsilon c}=\sqrt{-3/\Lambda}$, in the more convenient
form
\begin{eqnarray}
&&\d s^2={a^2\over t_1^2}\sinh^2 R_1\cos^2 \phi_1\,
 (\d t_1^2-\d\rho_1^2)-a^2(\d R_1^2+\sinh^2 R_1\d \phi_1^2) \nonumber\\
 &&\qquad\qquad -\sinh R_1\cos\phi_1\, H(R_1,\phi_1)
   \delta(t_1-\rho_1)(\d t_1-\d\rho_1)^2 \ .
  \label{E5.9}
 \end{eqnarray}
Again, it can be shown by direct calculations that (\ref{E5.9})
is just the solution (\ref{E4.1}) in coordinates introduced by the
parametrization
 \begin{eqnarray}
 Z_0 &=& -a\cosh R_1 +{a^2\over t_1}\sinh R_1\cos\phi_1 \nonumber\\
 Z_1 &=& \quad a\cosh R_1 -{a^2\over t_1}\sinh R_1\cos\phi_1
    +{\rho_1^2-t_1^2\over2t_1}\sinh R_1\cos\phi_1  \nonumber\\
 Z_2 &=& {\rho_1\over t_1}\>a\sinh R_1\cos\phi_1 \\
 Z_3 &=& \quad a\sinh R_1\sin\phi_1 \nonumber\\
 Z_4 &=& \quad a\cosh R_1 \hskip8pc
 +{\rho_1^2-t_1^2\over2t_1}\sinh R_1\cos\phi_1\ . \nonumber
 \label{E5.10}
 \end{eqnarray}

For the impulsive $PK(\Lambda^-)II$ subclass, the transformation
 \begin{equation}
 \xi = \sqrt{2} a\, e^{i\phi_2}\tanh{R_2\over 2},\qquad
   v = -{a^2\over \rho_2},\qquad
   u = t_2 - \rho_2,  \label{E5.11}
 \end{equation}
can be used to express the metric (\ref{E2.17}) in the form
\begin{eqnarray}
&&\d s^2={a^2\over \rho_2^2}\cosh^2 R_2\,
 (\d t_2^2-\d\rho_2^2)-a^2(\d R_2^2+\sinh^2 R_2\d \phi_2^2) \nonumber\\
 &&\qquad\qquad -\cosh R_2\, H(R_2,\phi_2)
   \delta(t_2-\rho_2)(\d t_2-\d\rho_2)^2 \ .
  \label{E5.12}
 \end{eqnarray}
However, this simple form also follows from (\ref{E4.1}) if we consider a
parametrization
 \begin{eqnarray}
 Z_0 &=& {\cosh R_2\over 2\rho_2}(t_2^2-\rho_2^2-a^2) \nonumber\\
 Z_1 &=& {\cosh R_2\over 2\rho_2}(t_2^2-\rho_2^2+a^2)  \nonumber\\
 Z_2 &=& a\sinh R_2\cos\phi_2 \\
 Z_3 &=& a\sinh R_2\sin\phi_2 \nonumber\\
 Z_4 &=& {t_2\over \rho_2}\, a\cosh R_2\ . \nonumber
 \label{E5.13}
 \end{eqnarray}

Finally, impulsive metric (\ref{E2.18}) of the $PK(\Lambda^-)III_K$
subclass for $H=\sqrt2\,H(\xi,\bar\xi)\delta(u)$ using a
transformation
 \begin{equation}
 \xi = \sqrt{2} a\, e^{i\phi_3}\tanh{R_3\over 2},\qquad
   v = {a\over \sqrt 2}(\eta-\chi),\qquad
   u = {a\over \sqrt 2}(\eta+\chi),  \label{E5.14}
 \end{equation}
reads
\begin{eqnarray}
&&\d s^2=a^2(\cosh R_3+\sinh R_3\cos \phi_3)^2\,
 (\d \eta^2-\d\chi^2)-a^2(\d R_3^2+\sinh^2 R_3\d \phi_3^2) \nonumber\\
 &&\qquad\qquad -a(\cosh R_3+\sinh R_3\cos \phi_3)\, H(R_3,\phi_3)
   \delta(\eta+\chi)(\d \eta+\d\chi)^2 \ ,
  \label{E5.15}
 \end{eqnarray}
which is exactly the metric of Eq. (21) in \cite{PodGri98} representing
the solution  (\ref{E4.1}) in global coordinate system given by
 \begin{eqnarray}
Z_0&=&a(\cosh R_3+\sinh R_3\cos\phi_3)\eta \nonumber \\
Z_1&=&a(\cosh R_3+\sinh R_3\cos\phi_3)\chi \nonumber \\
Z_2&=&a \sinh R_3\cos\phi_3
-{\textstyle{1\over2}}a(\cosh R_3+\sinh R_3\cos\phi_3)(\chi^2-\eta^2)
\label{E5.16} \\
Z_3&=&a\sinh R_3\sin\phi_3 \nonumber \\
Z_4&=&a\cosh R_3 \hskip2pc +{\textstyle{1\over2}}a(\cosh R_3+\sinh
  R_3\cos\phi_3)(\chi^2-\eta^2)\ . \nonumber
 \end{eqnarray}

We conclude this section by explicit transformations
relating the three different forms of impulsive $P(\Lambda)$
solutions for $\Lambda<0$ given above. The transformation between
the metric (\ref{E5.9}) of the $PK(\Lambda)I$ subclass for
$\Lambda<0$ and the metric (\ref{E5.12}) of the $PK(\Lambda^-)II$ subclass
is
 \begin{eqnarray}
 t_1^2     &=& t_2^2-\rho_2^2(1-\tanh^2 R_2\cos^2\phi_2)\ , \nonumber\\
 \rho_1    &=& \rho_2\tanh R_2\cos\phi_2 \ ,  \nonumber\\
 \cosh R_1 &=& {t_2\over \rho_2}\cosh R_2 \ , \\
 \sin\Phi_1&=& -{\sinh R_2\sin\phi_2\over
   \sqrt{t_2^2\rho_2^{-2}\cosh^2 R_2-1}} \ . \nonumber
 \label{E5.20}
 \end{eqnarray}
(Note that $t_1^2-\rho_1^2=t_2^2-\rho_2^2$ and
$\sinh R_1\sin\phi_1=-\sinh R_2\sin\phi_2$).
Performing the transformation
 \begin{eqnarray}
 \tilde\eta &=& {\cal D}{\cosh R_2\over 2\rho_2}(t_2^2-\rho_2^2-a^2)\ , \nonumber\\
 \tilde\chi &=& {\cal D}{\cosh R_2\over 2\rho_2}(t_2^2-\rho_2^2+a^2)\ ,  \nonumber\\
 x    &=& a{\cal D} \ , \\
 y    &=& a{\cal D}\sinh R_2\sin\phi_2 \ , \nonumber
 \label{E5.21}
 \end{eqnarray}
where ${\cal D}=(t_2\rho_2^{-1}\cosh R_2+\sinh R_2\cos \phi_2)^{-1}$,
the metric (\ref{E5.12}) of the $PK(\Lambda^-)II$ subclass goes
over to impulsive solution of the Siklos type
 \begin{equation}
 \d s^2= {a^2\over x^2}\left( \d\tilde\eta^2-\d\tilde\chi^2-\d x^2-\d y^2
-S(x,y)\delta(\tilde\eta+\tilde\chi)(\d\tilde\eta+\d\tilde\chi)^2
\right)\ .
 \label{E5.22}
 \end{equation}
This was shown in \cite{PodGri98} to be equivalent to the metric
(\ref{E5.15}) of the $PK(\Lambda^-)III_K$ class
through the transformation
 \begin{equation}
 \tilde\eta=a\eta, \quad \tilde\chi=a\chi, \quad
 x={a\over\cosh R_3+\sinh R_3\cos\phi_3}, \quad
 y={a\sinh R_3\sin\phi_3\over\cosh R_3+\sinh R_3\cos\phi_3}\ , \label{E5.23}
 \end{equation}
with an identification $S(x,y)=xH(x,y)/a$.

\section{Continuous coordinates for non-expanding impulsive
         gravitational waves}

\medskip\noindent
We have shown that all non-expanding impulsive waves of the
$P(\Lambda)$ class can be written either in the form (\ref{E2.14})
with $g=g(\xi)\delta(u)$, i.e. as {\it pp}-waves in Minkowski
spacetime, or in the form (\ref{E4.1}) in de~Sitter and anti-de~Sitter
backgrounds. Interestingly, both metrics can be expressed in a
unified form
 \begin{equation}
\d s^2= \d u\d r -\d \zeta\d\bar\zeta -\epsilon\d{Z_4}^2
  -H(\zeta,\bar\zeta)\delta(u)\d u^2  \ . \label{E6.1}
 \end{equation}
Indeed, the impulsive metric (\ref{E2.14}) arises
by setting $\epsilon=0$ and performing a transformation
 \begin{equation}
r=2v, \qquad
\zeta=\sqrt{2}\,\xi\ , \qquad
\bar\zeta=\sqrt{2}\,\bar\xi\ . \label{E6.2}
 \end{equation}
The relation to (\ref{E4.1}) follows from $\epsilon=\pm1$
and a transformation
 \begin{equation}
u=Z_0+Z_1, \qquad r=Z_0-Z_1, \qquad
\zeta=Z_2+iZ_3, \qquad \bar\zeta=Z_2-iZ_3\ ; \label{E6.3}
 \end{equation}
note that in this case we consider the dependence of $H(Z_2, Z_3, Z_4)$
on the argument $Z_4$ given by $Z_4=\pm\sqrt{a^2-\epsilon\zeta\bar\zeta}$
which follows from (\ref{E4.3}).

Now, performing a coordinate transformation
 \begin{eqnarray}
\zeta&=&\eta+u\Theta(u)H_{,\bar\eta}\ , \nonumber \\
r  &=& w+\Theta(u) H + u\Theta(u) H_{,\eta}H_{,\bar\eta}\ , \label{E6.4}
 \end{eqnarray}
where $\Theta(u)$ is the Heaviside step function
($\Theta=0$ for $u<0$, $\Theta=1$ for $u>0$)
and $H(\eta,\bar\eta)$ is the same function as in (\ref{E6.1}) but expressed
in terms of the new coordinate $\eta$ (note that $\eta=\zeta$ at $u=0$) we get
 \begin{equation}
\d s^2= \d u\d w -\epsilon\d{Z_4}^2
  -\left|\d \bar\eta+u\Theta(u)(H_{,\eta\eta}\d\eta
      +H_{,\eta\bar\eta}\d\bar\eta)\right|^2  \ . \label{E6.5}
 \end{equation}
This form of non-expanding impulsive waves in
Minkowski, de~Sitter and anti-de~Sitter space-times is explicitly
continuous for all values of $u$ (delta functions appear only in the
components of the curvature tensor).

In a Minkowski background, the cosmological constant $\Lambda$
vanishes, $\epsilon=0$ and
$H(\zeta,\bar\zeta)=2[g(\zeta/\sqrt2)+\bar g(\bar\zeta\sqrt2)]$ so that
$H_{,\eta\eta}=g''$, $H_{,\eta\bar\eta}=0$; then the metric
(\ref{E6.5}) reduces to
 \begin{equation}
\d s^2= \d u\d w -\left|\d \bar\eta
   +u\Theta(u)g''(\eta)\d\eta\right|^2\ . \label{E6.6}
 \end{equation}
This continuous system for a general impulsive {\it pp}-wave
was found in \cite{AB} and independently in \cite{PodVes98}.
In particular, when $g(\zeta)=C\zeta^n$
where $C=|C|\exp(in\varphi_n)$ is an arbitrary complex constant
and $n$ is an integer,
it is convenient to introduce two real coordinates
$\rho,\varphi$ by
$\eta=\rho\exp[i(\varphi-\varphi_n)]$.
In these coordinates the metric (\ref{E6.6}) takes the form
\begin{eqnarray}
\d s^2 &=& \d u\d w
-[1-u\Theta(u)D\rho^{n-2}]^2(\d\rho^2+\rho^2\d\varphi^2)
    \nonumber\\
   && -4u\Theta(u)D\rho^{n-2}
            [\cos({\textstyle \frac{n}{2}}\varphi)\d\rho
            -\sin({\textstyle \frac{n}{2}}\varphi)\rho\d\varphi]^2
        \ ,        \label{E6.7}
\end{eqnarray}
where $D=n(n-1)|C|$. We may also  assume coordinates
$x=\rho\cos\varphi$, $y=\rho\sin\varphi$,
\begin{eqnarray}
\d s^2&=&  \d u\d w + 4DB_n u\Theta(u)\, \d x\d y \nonumber\\
&&-[1+2DA_n u\Theta(u)+D^2(x^2+y^2)^{n-2}u^2\Theta(u)]\,\d x^2
                          \label{E6.8}\\
      &&  -[1-2DA_n u\Theta(u)+D^2(x^2+y^2)^{n-2}u^2\Theta(u)]\,\d y^2
 \nonumber
\end{eqnarray}
where $A_n(x,y)={\cal R}e\,\{(x+iy)^{n-2}\}$,
$B_n(x,y)={\cal I}m\,\{(x+iy)^{n-2}\}$.
In particular, the metric (\ref{E6.7}) with $n=0$, $D\not=0$
reduces to the continuous from of the Aichelburg and Sexl solution
\cite{AicSex71} presented in \cite{DE78}, \cite{DEP93}. For $n=2$
the metric (\ref{E6.8}) gives the well-known form of the
homogeneous impulsive {\it pp}-wave \cite{Penrose}.

Continuous coordinate system for non-expanding impulsive waves
in de~Sitter and anti-de~Sitter backgrounds is given by
(\ref{E6.5}) with $\epsilon=+1$ for $\Lambda>0$ and
$\epsilon=-1$ for $\Lambda<0$. The function $H(\eta, \bar\eta)$
for first few multipole terms $H_m$ given by (\ref{E4.5}) and
(\ref{E4.6}) is
 \begin{eqnarray}
H_0(\eta,\bar\eta)&=&\pm
{1\over2}\sqrt{1-\epsilon a^{-2}\eta\bar\eta}
  \,\log\left|{1\pm \sqrt{1-\epsilon a^{-2}\eta\bar\eta}
        \over1\mp \sqrt{1-\epsilon a^{-2}\eta\bar\eta}}\right|-1, \nonumber \\
H_1(\eta,\bar\eta)&=&\left(-{\epsilon\eta\bar\eta\over4a}
  \,\log\left|{1\pm \sqrt{1-\epsilon a^{-2}\eta\bar\eta}
        \over1\mp \sqrt{1-\epsilon a^{-2}\eta\bar\eta}}\right|
   \mp\frac{a}{2}\sqrt{1-\epsilon a^{-2}\eta\bar\eta} \right)
  \left(\frac{e^{i\phi_1}}{\eta}+\frac{e^{-i\phi_1}}{\bar\eta}\right),
                                     \nonumber \\
H_2(\eta,\bar\eta)&=& \frac{a^2}{\eta^2}e^{i2\phi_2}
    +\frac{a^2}{{\bar\eta}^2}e^{-i2\phi_2}    , \label{E6.9}\\
H_3(\eta,\bar\eta)&=&\mp 4\sqrt{1-\epsilon a^{-2}\eta\bar\eta}
  \left(\frac{a^3}{\eta^3}e^{i3\phi_3}
    +\frac{a^3}{{\bar\eta}^3}e^{-i3\phi_3}\right).   \nonumber
 \end{eqnarray}
These functions satisfy the vacuum field equation
$(\eta^2 H_{,\eta})_{,\eta}+({\bar\eta}^2 H_{,\bar\eta})_{,\bar\eta}
 +2(\eta\bar\eta-2\epsilon a^2)H_{,\eta\bar\eta}-2H = 0$
everywhere except for the point sources at $\eta=0$.
Substituting these particular forms into (\ref{E6.5}) we get
continuous coordinates for the first multipole terms.
Note that the simplest form of the metric arises for the quadruple
term $H_2$,
 \begin{equation}
\d s^2= \d u\d w -\epsilon\d{Z_4}^2
  -\left|\d
\bar\eta+Cu\Theta(u)\frac{\d\eta}{\eta^4}
      \right|^2  \ , \label{E6.10}
 \end{equation}
where $C=6a^2e^{i2\phi_1}$.

The continuous form (\ref{E6.5}) of a general impulsive solution
 for non-vanishing $\Lambda$ is explicit and simple.
However, it is written in the five-dimensional notation
so that we must always restrict the metric on the hyperboloid (\ref{E4.2})
which in our  coordinates is given by
 $ uw -\eta\bar\eta-\epsilon{Z_4}^2 =-\epsilon a^2$ for $u<0$.
It may thus be useful to find a continuous metric also in standard
four-dimensional form. There are many suitable
parametrizations of the de Sitter hyperboloid (for $\epsilon=1$), cf.
\cite{PodGri97}. We can consider, for example, global coordinates
($u,x,y,z$)
such that
 \begin{equation}
 w   = {x^2+y^2-a^2\over u}+{uz^2\over\ a^2},\qquad
 Z_4 = {uz\over a},\qquad  \eta= x+iy,  \label{E6.11}
 \end{equation}
in which the metric (\ref{E6.5}) takes the form
 \begin{eqnarray}
\d s^2&=&{a^2\over u^2}\,\d u^2-{u^2\over a^2}\,\d z^2
  -\left(\d x-{x\over u}\d u\right)^2
  -\left(\d y-{y\over u}\d u\right)^2 \nonumber\\
 &&\qquad -u\Theta(u)\,
   [(C+A)\,\d x^2 + (C-A)\,\d y^2 + 2B\,\d x\d y]  \ ,
  \label{E6.12}
 \end{eqnarray}
where
$A=(1+uH_{,\eta\bar\eta})\,(H_{,\eta\eta}+H_{,\bar\eta\bar\eta})$,
$B=(1+uH_{,\eta\bar\eta})i(H_{,\eta\eta}-H_{,\bar\eta\bar\eta})$
and $C=2 H_{,\eta\bar\eta}+
 u(H_{,\eta\eta}H_{,\bar\eta\bar\eta}-H_{,\eta\bar\eta}^2)$.
The function $H$ and all its derivatives are now
expressed in terms of the coordinates $x$ and $y$ so that the
coefficients $A$, $B$ and $C$ are independent of $z$.
A simple transformation $\psi=a^2/u$,
$\tilde x=ax/u$ and $\tilde y=ay/u$ reveals for $u<0$
a standard conformally flat metric of the de Sitter universe,
$\d s^2=a^2\psi^{-2}\,(\d\psi^2-\d\tilde x^2-\d\tilde y^2-\d z^2)$.
Note that the singularity at $\psi=\infty$ is only a coordinate one.

In order to find a suitable four-dimensional system
for non-expanding impulsive waves in the anti-de Sitter background
($\Lambda<0$, $\epsilon=-1$) it is more convenient to start from the
Siklos form of the metric (\ref{E5.22}) which is explicitly
conformal to {\it pp}-waves. By a simple reparametrization
 \begin{equation}
u=\tilde\eta+\tilde\chi, \qquad r=\tilde\eta-\tilde\chi, \qquad
\zeta=x+iy, \qquad \bar\zeta=x-iy, \label{E6.14}
 \end{equation}
the metric (\ref{E5.22}) goes over to
 \begin{equation}
 \d s^2= {4a^2\over (\zeta+\bar\zeta)^2}\left( \d u\d r
  -\d\zeta\d\bar\zeta-S(\zeta,\bar\zeta)\delta(u)\d u^2 \right)\ .
   \label{E6.15}
 \end{equation}
We may use the transformation (\ref{E6.4}) with $H=S$ and we get
a continuous form
 \begin{equation}
\d s^2=\frac{4a^2}
 {\left[\eta+\bar\eta+u\Theta(u)(S_{,\eta}+S_{,\bar\eta})\right]^2}
\left(\d u\d w-\left|\d \bar\eta+u\Theta(u)(S_{,\eta\eta}\d\eta
      +S_{,\eta\bar\eta}\d\bar\eta)\right|^2\right)  \ . \label{E6.16}
 \end{equation}
Note that in his paper \cite{Siklos85}, Siklos has shown that,
for the vacuum case (except for some possible point sources),
the general solution for $S$ is of the form
$ S(\eta,\bar\eta)=\textstyle{\frac{1}{2}}(\eta+\bar\eta)
  (f_{,\eta} +\bar f_{,\bar\eta})-(f+\bar f)$  where $f$
is an arbitrary function of $\eta$. In particular, the Kaigorodov
solution (which is the simplest representative of the Siklos
class, see \cite{Podol98}) is given by $f=\textstyle{\frac{1}{4}}\eta^3$
so that the continuous system for an impulsive Kaigorodov spacetime is
\begin{equation}
\d s^2=\frac{a^2}{X^2}(1+\textstyle{\frac{3}{2}}u\Theta(u)X)^{-2}
\left[\d u\d w-(1+3u\Theta(u)X)^2 \d X^2 - \d Y^2 \right]  \ ,
\label{E6.17}
\end{equation}
where we introduced $\eta=X+iY$.

\section*{Acknowledgments}

This work was supported by the grant GACR-202/96/0206 of the Czech Republic
and the grant GAUK-230/96 of the Charles University.

\end{document}